\documentstyle[preprint,revtex,epsf]{aps}

\begin{document}
\begin{title}
The Rare Decay $B\rightarrow K^{\ast}\gamma$: A More Precise
Calculation.
\end{title}
\author{Patrick J. O'Donnell}
\begin{instit}
Department of Physics,
University of Toronto,\\
60 St. George Street,
Toronto, Canada M5S 1A7
\end{instit}
\vspace{.1in}
\centerline{and}
\vspace{.1in}
\author{Humphrey K.K. Tung}
\begin{instit}
Institute of Physics,
Academia Sinica,\\
Taipei, Taiwan 11529, R.O.C.
\end{instit}
\vspace{.2in}
\centerline{\today}
\vspace{.2in}
\begin{abstract}
Efforts to predict the rare exclusive decay
$B\rightarrow K^{\ast}\gamma$ from the well known inclusive decay
$b\rightarrow s\gamma$ are frustrated by the effect of the large
recoil momentum. We show how to reduce the large uncertainty in
calculating this decay by relating $B\rightarrow K^{\ast}\gamma$ to the
semileptonic process $B\rightarrow\rho e\bar{\nu}$ using the
heavy-quark symmetry in $B$ decays and $SU(3)$ flavor symmetry.
A direct measurement of the $q^{2}$-spectrum for the
semileptonic decay can provide accurate information
for the exclusive rare decay.
\end{abstract}

\vspace{0.5in}
\noindent
{\bf UTPT-93-02}\\
{\bf IP-ASTP-14-92}

\newpage
The inclusive rare decay $B\rightarrow X_{s}\gamma$ is now well
understood in the context of the standard model \cite{stan} and the
experimental upper bound \cite{CLEO1} of $8.4\times 10^{-4}$ for the branching
ratio is already playing an important role \cite{Hew} in constraining the
parameters of models other than the standard model. On the other hand, the
most likely experimental observation to be made will be the exclusive
decay $B\rightarrow K^{\ast}\gamma$. The recent limit from the CLEO
collaboration \cite{CLEO2} of the branching ratio for this mode is
$0.92\times 10^{-4}$.  It is this
exclusive rare decay $B\rightarrow K^{\ast}\gamma$, however, which
is the least well known theoretically due to the large
recoil momentum of the $K^{\ast}$ meson \cite{OT}.
A recent paper \cite{BD} points out that heavy-quark symmetry together with
$SU(3)$ flavor symmetry could relate the rare decay
$B\rightarrow K^{\ast}\gamma$ to a measurement of
the semileptonic decay $B\rightarrow\rho e\bar{\nu}$.
However, the relation is only valid at a single point
in the Dalitz plot, a point where the semileptonic decay vanishes,
so that there would still be a large uncertainty in such a measurement.

In this paper we obtain a similar relation that relates the exclusive
rare decay $B\rightarrow K^{\ast}\gamma$
to the spectrum in $q^{2}$ for the semileptonic decay
$B\rightarrow\rho e\bar{\nu}$.
The $q^{2}$-spectrum for $B\rightarrow\rho e\bar{\nu}$ does not vanish
at $q^{2}=0$ and so a direct measurement of the spectrum at this point can
provide accurate information for $B\rightarrow K^{\ast}\gamma$. Of
necessity this new result requires an extension of the heavy-quark
symmetries to a consideration of the $K^{\ast}$ and $\rho$ systems.
We show that this is not the same as demanding the $K^{\ast}$
or $\rho$ to be a heavy-quark system in the conventional sense.
Our result dramatically reduces the uncertainty from the earlier
calculations.

First we discuss the application of the heavy-quark symmetry. Usually,
it is the hadronic systems with a $b$ or a $c$ quark that have these
symmetries. Here we derive the relations for matrix elements
of either a $B$ or a $B^{\ast}$ with an unspecified vector meson $V$.
We show how to extend the heavy-quark symmetry relations
to the case when the meson $V$ is the $K^{\ast}$ or $\rho$,
and we estimate the possible errors using a set of quark-model
calculations, both for nonrelativistic and for relativistic
cases.  Then we use the results to give a reliable
relation between the decay $B\rightarrow K^{\ast}\gamma$ and
the $q^{2}$-spectrum for $B\rightarrow\rho e\bar{\nu}$.

\vspace{.3in}
\centerline{{\bf I}. Heavy-Quark Symmetry Relations}

We first recapitulate the derivation of the heavy-quark symmetry
relations for $B$ decay.
The hadronic matrix elements relevant to the decay
$B(b\bar{q})\rightarrow V(Q\bar{q})$ are given by
\begin{eqnarray}
\langle V(k,\epsilon)|\bar{Q}\gamma_{\mu}b|B(p_{B})\rangle &=&
2T_{1}(q^{2})i\varepsilon_{\mu\nu\lambda\sigma}\epsilon^{\ast\nu}
p_{B}^{\lambda}k^{\sigma}\,\,\, ,\\
\langle V(k,\epsilon)|\bar{Q}\gamma_{\mu}\gamma_{5}b|B(p_{B})\rangle
&=& -2(m^{2}_{B}-m^{2}_{V})T_{2}(q^{2})\epsilon^{\ast}_{\mu}
-2T_{3}(q^{2})(\epsilon^{\ast}\cdot q)(p_{B}+k)_{\mu}\nonumber\\
&& -2T_{4}(q^{2})(\epsilon^{\ast}\cdot q)(p_{B}-k)_{\mu}\,\,\, ,\\
\langle V(k,\epsilon)|\bar{Q}i\sigma_{\mu\nu}q^{\nu}b_{R}|B(p_{B})\rangle
&=& f_{1}(q^{2})i\varepsilon_{\mu\nu\lambda\sigma}
\epsilon^{\ast\nu}p^{\lambda}_{B} k^{\sigma}\nonumber\\
&& + \left[(m^{2}_{B}-m^{2}_{V})\epsilon^{\ast}_{\mu}-
(\epsilon^{\ast}\cdot q)(p_{B}+k)_{\mu}\right]f_{2}(q^{2})\nonumber\\
&& + (\epsilon^{\ast}\cdot q)\left[(p_{B}-k)_{\mu}-\frac{q^{2}}{(m^{2}_{B}
-m^{2}_{V})}(p_{B}+k)_{\mu}\right]f_{3}(q^{2})\,\,\, ,\nonumber\\
\end{eqnarray}
where $q=p_{B}-k$. We show below that the hadronic form factors
$f_{1,2,3}(q^{2})$ and $T_{1,2,3,4}(q^{2})$ for the decay
$B\rightarrow V$ can all be related using just
the spin symmetry and static limit of the heavy $b$ quark.

In the heavy $b$ limit, the spin of the $b$ quark is
decoupled from all other light fields in $B$ \cite{IW1}.
We can therefore construct the spin operator $S^{Z}_{b}$ for the $b$
quark such that
\[
S^{Z}_{b}|B(b\bar{q})\rangle
= \frac{1}{2}|B^{\ast}_{l}(b\bar{q})\rangle\,\,\, ,
\,\,\, S^{Z}_{b}|B^{\ast}_{l}(b\bar{q})\rangle
= \frac{1}{2}|B(b\bar{q})\rangle\,\,\, ,
\]
where $B^{\ast}_{l}$ stands for a longitudinal vector $B^{\ast}$ meson.
In $|B\rangle$ and $|B^{\ast}_{l}\rangle$,
the spatial momentum of the $b$ quark is in
the $z$-direction for the $b$ spinor to be an eigenstate
of $S^{Z}_{b}$. Using the relation
$\langle V|\,\bar{Q}\Gamma b\,|B\rangle =
-2\langle V|\,[S^{Z}_{b},\bar{Q}\Gamma b]\,|B^{\ast}_{l}\rangle $
for $\Gamma$ any product of $\gamma$ matrices,
we have the following identities between the
$B\rightarrow V$ and $B^{\ast}_{l}\rightarrow V$ matrix elements:
\begin{eqnarray}
\langle V|A_{0}|B\rangle &=& -\langle V|V_{3}|B^{\ast}_{l}
\rangle\,\,\, ,\label{v1}\\
\langle V|A_{3}|B\rangle &=& -\langle V|V_{0}|B^{\ast}_{l}
\rangle\,\,\, ,\label{v2}\\
\langle V|V_{\pm}|B\rangle &=& \mp\langle V|V_{\pm}|B^{\ast}_{l}
\rangle\,\,\, ,\label{v3}\\
\langle V|V_{0}|B\rangle &=& -\langle V|A_{3}|B^{\ast}_{l}
\rangle\,\,\, ,\label{a1}\\
\langle V|V_{3}|B\rangle &=& -\langle V|A_{0}|B^{\ast}_{l}
\rangle\,\,\, ,\label{a2}\\
\langle V|A_{\pm}|B\rangle &=& \mp\langle V|A_{\pm}|B^{\ast}_{l}
\rangle\,\,\, ,\label{a3}
\end{eqnarray}
where $V_{\mu}=\bar{Q}\gamma_{\mu}b$ and
$A_{\mu}=\bar{Q}\gamma_{\mu}\gamma_{5}b$.
The covariant expansions of the vector and axial-vector matrix elements
for the decay $B^{\ast}(b\bar{q})\rightarrow V(Q\bar{q})$ are defined
to be
\begin{eqnarray}
\langle V(k,\epsilon)|\bar{Q}\gamma_{\mu}b|B^{\ast}(p_{B},\zeta)\rangle
&=& \left[\,(\zeta\cdot\epsilon^{\ast}) {\cal A}_{1}(q^{2}) + (\zeta\cdot q)
(\epsilon^{\ast}\cdot q){\cal A}_{2} (q^{2})\,\right] (p_{B}+k)_{\mu}
\nonumber\\
&& + \left[\,(\zeta\cdot\epsilon^{\ast}) {\cal B}_{1}(q^{2}) + (\zeta\cdot q)
(\epsilon^{\ast}\cdot q){\cal B}_{2} (q^{2})\,\right] (p_{B}-k)_{\mu}
\nonumber\\
&& + {\cal C}(q^{2}) (\epsilon^{\ast}\cdot q) \zeta_{\mu}
+ {\cal D}(q^{2}) (\zeta\cdot q) \epsilon^{\ast}_{\mu}
\,\,\, ,\\
\langle V(k,\epsilon)|\bar{Q}\gamma_{\mu}\gamma_{5}b
|B^{\ast}(p_{B},\zeta)\rangle
&=& {\cal E}(q^{2})i\varepsilon_{\mu\nu\lambda\sigma}\epsilon^{\ast\nu}
\zeta^{\lambda}(p_{B}+k)^{\sigma}
+ {\cal F}(q^{2})i\varepsilon_{\mu\nu\lambda\sigma}\epsilon^{\ast\nu}
\zeta^{\lambda}(p_{B}-k)^{\sigma} ,\nonumber\\
\end{eqnarray}
where $\zeta$ and $\epsilon^{\ast}$ are the polarization vectors
of $B^{\ast}$ and $V$, respectively.
Using the matrix identities in Eqs. (\ref{v1})-(\ref{a3}),
we can relate $T_{1,2,3,4}$ to the $B^{\ast}\rightarrow V$ form factors:-
\begin{eqnarray}
2m_{B}T_{1}&=& ({\cal A}_{1}-{\cal B}_{1})\,\,\, ,\nonumber\\
2(m^{2}_{B}-m^{2}_{V})T_{2}&=& m_{B}({\cal A}_{1}+{\cal B}_{1})
+ E_{V}({\cal A}_{1}-{\cal B}_{1})\,\,\, ,\nonumber\\
2m_{B}(T_{3}-T_{4})&=&-({\cal A}_{1}-{\cal B}_{1})\,\,\, , \nonumber\\
2m_{B}(T_{3}+T_{4})&=& -({\cal A}_{1}+{\cal B}_{1})-{\cal C}
\,\,\, ,\nonumber\\
{\cal D}&=& ({\cal A}_{1}-{\cal B}_{1})\,\,\, ,\nonumber\\
{\cal E}&=& {\cal A}_{1}\,\,\, ,\nonumber\\
{\cal F}&=& {\cal B}_{1}\,\,\, ,\nonumber\\
{\cal A}_{2}&=& {\cal B}_{2}=0\,\,\, .\label{relation}
\end{eqnarray}
Since the spatial momentum of the $b$ quark is defined in
the $z$-direction, the above relations are worked out in the $B$
rest frame. We choose the longitudinal polarization
vector for $B^{\ast}_{l}$ to be $\zeta^{\mu}_{l}=(0;0,0,1)$
and define the momentum of $V$ to be
$k^{\mu}=(E_{V};k^{1},k^{2},k^{3})$,
where $E_{V}=(m^{2}_{B}+m^{2}_{V}-q^{2})/(2m_{B})$.
The resulting form-factor relations in Eq. (\ref{relation}) are
consistent with those in Ref. \cite{IW1} using the
spin symmetry of
a heavy $Q$, except for the relation $T_{3}+T_{4}=0$, which is
missing here.

We can relate the form factors $f_{1,2,3}$ to the form factors $T_{1,2,3,4}$
using the static limit of the $b$ quark. In the $B$ rest frame,
the static $b$-quark spinor satisfies the equation of motion $\gamma_{0}b=b$.
We then have the relations between the $\gamma_{\mu}$
and $\sigma_{\mu\nu}$ matrix elements \cite{IW2}:-
\begin{eqnarray}
\langle V | \bar{Q}\gamma_{i}b|B\rangle &=&
\langle V | \bar{Q}i\sigma_{0i}b|B\rangle\,\,\, ,
\\
\langle V | \bar{Q}\gamma_{i}\gamma_{5}b|B\rangle
&=& - \langle V | \bar{Q}i\sigma_{0i}\gamma_{5}b|B
\rangle\,\,\, .
\end{eqnarray}
This gives the form-factor relations
\begin{eqnarray}
f_{1}&=& -(m_{B}-E_{V})T_{1}-\frac{(m^{2}_{B}-m^{2}_{V})}{m_{B}}T_{2}
\,\,\, ,\nonumber\\
f_{2}&=& -\frac{1}{2}\left[ (m_{B}-E_{V})-(m_{B}+E_{V})
\frac{q^{2}}{m^{2}_{B}-m^{2}_{V}}\right]T_{1}
-\frac{1}{2m_{B}}\left( m^{2}_{B}-m^{2}_{V}+q^{2}\right) T_{2}
\,\,\, ,\nonumber\\
f_{3}&=& -\frac{1}{2}(m_{B}+E_{V})T_{1}+\frac{1}{2m_{B}}
(m^{2}_{B}-m^{2}_{V})(T_{1}+T_{2}+T_{3}-T_{4})\,\,\, .
\end{eqnarray}
Using the spin-symmetry relations in Eq. (\ref{relation}) , we can also
write $f_{1,2,3}$ in terms of the $B^{\ast}\rightarrow V$ form factors as,
\begin{eqnarray}
f_{1}&=& -{\cal A}_{1}\,\,\, ,\nonumber\\
f_{2}&=& -\frac{1}{2}{\cal A}_{1}-\frac{1}{2}
\left( \frac{q^{2}}{m^{2}_{B}-m^{2}_{V}}\right) {\cal B}_{1}
\,\,\, ,\nonumber\\
f_{3}&=& \frac{1}{2}{\cal B}_{1}\,\,\, .
\end{eqnarray}

Thus, using only the spin symmetry and static limit of the heavy
$b$ quark, we can relate the $B\rightarrow V$ hadronic form factors as,
\begin{eqnarray}
2(m^{2}_{B}-m^{2}_{V})T_{2}&=& [\,(m_{B}+m_{V})^{2}-q^{2}\, ]\,T_{1}
+ \alpha \,\,\, ,\nonumber\\
2T_{3}&=& -T_{1}+\frac{\beta}{2m_{B}}\,\,\, ,\nonumber\\
2T_{4}&=& T_{1}+\frac{\beta}{2m_{B}}\,\,\, ,\nonumber\\
f_{1}&=& -(m_{B}+m_{V})T_{1}-\frac{\alpha}{2m_{B}}\,\,\, ,\nonumber\\
2f_{2}&=& -\left[ \,(m_{B}+m_{V})^{2}-q^{2}\,\right]
\frac{T_{1}}{(m_{B}+m_{V})} - \frac{\alpha}{2m_{B}}\left(
1+\frac{q^{2}}{m^{2}_{B}-m^{2}_{V}}\right)\,\,\, ,\nonumber\\
2f_{3}&=& -(m_{B}-m_{V})T_{1}+\frac{\alpha}{2m_{B}}\,\,\, ,
\label{final}
\end{eqnarray}
where
\begin{eqnarray}
\alpha&=& m_{B}({\cal A}_{1}+{\cal B}_{1})
- m_{V}({\cal A}_{1}-{\cal B}_{1})\,\,\, ,\nonumber\\
\beta&=& -{\cal C}-({\cal A}_{1}+{\cal B}_{1})\,\,\, .\nonumber
\end{eqnarray}

If we ignore the $\alpha$ and $\beta$ terms in Eq. (\ref{final}),
the resulting form-factor relations are exactly the heavy-quark
symmetry relations obtained using the large-mass limits of both
$b$ and $Q$ quarks \cite{OT,IW1,IW2}.
The function $-\sqrt{4m_{B}m_{V}}T_{1}$ in this limit resembles
the role of the Isgur-Wise function, with absolute
normalization at $q^{2}=t_{m}$ given in the quark model as \cite{OT}
\[
-\sqrt{4m_{B}m_{V}}T_{1}(t_{m})=\left(\frac{2\beta_{B}\beta_{V}}
{\beta^{2}_{B}+\beta^{2}_{V}}\right)^{3/2} \approx 1\,\,\, ,
\]
where $\beta_{B}$ and $\beta_{V}$ are variational parameters
of the momentum wave functions for $B$ and $V$ respectively.
We shall show below that the $\alpha$ and $\beta$
terms in Eq. (\ref{final}) can be regarded as small corrections
to the heavy-quark relations coming from the weak binding, or
$\Lambda_{QCD}$ effects. Thus, the symmetry relations are
dominated by the $T_{1}$ terms.

To show that the RHS of Eq. (\ref{final}) is dominated by the
$T_{1}$ terms, we use the equation of motion for the $b$ quark
to get the matrix relation
$\, \langle V|\bar{Q}p\!\!\! /_{B} b|B^{\ast}\rangle
= m_{B}\langle V|\bar{Q}b|B^{\ast}\rangle \,$ and work in the
$V$ rest frame. If we assume the static limit of $Q$ so
that the equation of motion  $\bar{Q}=\bar{Q}\gamma_{0}$ is satisfied
we can relate the matrix elements for $\bar{Q}b$ and
$\bar{Q}\gamma_{0}b$ as,
\begin{equation}
\langle V|\bar{Q}p\!\!\! /_{B} b|B^{\ast}\rangle
\approx m_{B}\langle V|\bar{Q}\gamma_{0}b|B^{\ast}\rangle \,\,\, .
\label{motion}
\end{equation}
The additional form-factor
relations that follow from Eq. (\ref{motion}) are given by
\begin{equation}
-{\cal C} = ({\cal A}_{1}+{\cal B}_{1}) = \frac{m_{V}}{m_{B}}
({\cal A}_{1}-{\cal B}_{1})\,\,\, ,
\label{extra}
\end{equation}
corresponding to $\alpha = \beta = 0$ in Eq. (\ref{final}).
The correction to the static $Q$ assumption is proportional
to ${\bf p}/m_{Q}$, where
${\bf p}$ is the spatial momentum of $Q$ in the $V$ rest frame.
Since
\[
\,\, m_{V}=\sqrt{{\bf p}^{2}+m^{2}_{Q}} +
\sqrt{{\bf p}^{2}+m^{2}_{q}} + {\em Binding\,\,\, Energy}\,\,\, ,
\]
and $m_{V}\approx m_{Q}+m_{q}$ for a heavy enough vector meson, in the weak
binding
limit of $V$ it is easy to show that ${\bf p}/m_{Q}\ll 1$. The
corrections to Eq. (\ref{extra}) arising from the static $Q$ assumption
are therefore small, and consequently we have the suppression of
$\,\alpha, \beta\sim {\cal O}({\bf p}/m_{Q}) \,\, $ in this limit.

We can use the quark model to show explicitly the suppression of $\alpha$ and
$\beta$ in the weak binding limit.
In the quark-model calculations of $\alpha$, $\beta$, and $T_{1}$,
we have in the $B$ rest frame
\begin{eqnarray}
m_{B}\beta &=& -\alpha = \sqrt{4m_{B}E_{V}}
\left( h_{1}-\frac{h_{2}}{(E_{V}-m_{V})} \right)\,\,\, ,\label{beta}\\
T_{1} &=& -\sqrt{\frac{E_{V}}{m_{B}}}
\frac{h_{2}}{(E_{V}+m_{V})(E_{V}-m_{V})}\,\,\, ,\label{T1}
\end{eqnarray}
where
\begin{eqnarray}
h_{1} &=& \int d{\bf p}\,\phi_{V}\phi_{B}\,
\sqrt{\frac{E_{Q}+m_{Q}}{2E_{Q}}}
\sqrt{\frac{E_{b}+m_{b}}{2E_{b}}}\,
\left( 1+\frac{({\bf k}+{\bf p})\cdot {\bf p}}{(E_{Q}+m_{Q})(E_{b}+m_{b})}
\right)\,\,\, , \nonumber\\
h_{2} &=& \int d{\bf p}\,\phi_{V}\phi_{B}\,
\sqrt{\frac{E_{Q}+m_{Q}}{2E_{Q}}}
\sqrt{\frac{E_{b}+m_{b}}{2E_{b}}}\,
\left( \frac{{\bf k}\cdot{\bf p}}{(E_{b}+m_{b})}
+ \frac{{\bf k}\cdot({\bf k}+{\bf p})}{(E_{Q}+m_{Q})}\right)
\,\,\, . \label{model}
\end{eqnarray}
The term ${\bf k}$ is the recoil momentum of $V$ in the $B$ rest frame.
The energies of the $b$ and $Q$ quarks in Eq. (\ref{model}) are given by
$\,E_{b}=\sqrt{ {\bf p}^{2}+m^{2}_{b}}\,$ and
$\,E_{Q}=\sqrt{ ({\bf k}+{\bf p})^{2}+m^{2}_{Q}}\,$.
The terms $\phi_{B}({\bf p})$ and
$\phi_{V}({\bf p}+r{\bf k})$, where $r=m_{q}/(m_{Q}+m_{q})$, are the momentum
wave functions of $B$ and $V$, respectively.
In the weak binding limit of $V$ and with $m_{V}\approx m_{Q}+m_{q}$,
we have \cite{OT}
\begin{equation}
\frac{{\bf k}\cdot{\bf p}}{(E_{b}+m_{b})}
+ \frac{{\bf k}\cdot({\bf k}+{\bf p})}{(E_{Q}+m_{Q})}
\approx (E_{V}-m_{V})
\left( 1+\frac{({\bf k}+{\bf p})\cdot {\bf p}}{(E_{Q}+m_{Q})(E_{b}+m_{b})}
\right) \,\,\, .\label{weak}
\end{equation}
This gives in Eq. (\ref{model}) the relation
\begin{equation}
(E_{V}-m_{V})h_{1}\approx h_{2}\,\,\, ,\label{limit}
\end{equation}
which is insensitive to
the problem of how the overlap occurs between $\phi_{B}$ and $\phi_{V}$.
The suppression of $\alpha$ and $\beta$
in this limit are then obvious from Eq. (\ref{beta}).

In the quark model, the relation between $-\alpha$ (and $m_{B}\beta$) and
$T_{1}$ can also be written
\begin{eqnarray}
m_{B}\beta &=& - \alpha =
[\,(m_{B}+m_{V})^{2}-q^{2}\,]\, T_{1}\,\varepsilon \,\,\, ,
\label{suppression}
\end{eqnarray}
where $\varepsilon = 1 - (E_{V}-m_{V})h_{1}/h_{2}$.
The correction to Eq. (\ref{weak}) is given by
\begin{eqnarray}
\delta &=&
\left[ \,(m_{b}-m_{Q})-(E_{b}-E_{Q})-(E_{V}-m_{V})\,\right]
\left( 1+\frac{({\bf k}+{\bf p})\cdot {\bf p}}{(E_{Q}+m_{Q})(E_{b}+m_{b})}
\right) \nonumber\\
&& -2(m_{b}-m_{Q})
\frac{({\bf k}+{\bf p})\cdot {\bf p}}{(E_{Q}+m_{Q})(E_{b}+m_{b})}
\,\,\, ,\label{delta}
\end{eqnarray}
which corresponds to the binding effects in $V$. From Eq. (\ref{delta}),
it is then easy to show that $\varepsilon\approx 0$ in the weak
binding limit of $V$ and with $m_{V}\approx m_{Q}+m_{q}$.

Using the Gaussian wave functions for $\phi_{B}$
and $\phi_{V}$ \cite{OT,ISGW}, we obtain throughout the whole kinematic
region a rather stable ratio for $(E_{V}-m_{V})h_{1}/h_{2}$, which is
between $0.95$ and $1.04$ for $B\rightarrow K^{\ast}$, and between
$0.89$ and $0.98$ for
$B\rightarrow\rho$. These two ranges include
the large uncertainty in the quark model associated with
different recoil
dependencies in the wave function overlap \cite{OT}. In all cases
the value for $\varepsilon$ is very small with
$|\varepsilon|<0.05$ for $B\rightarrow K^{\ast}$ and
$|\varepsilon|<0.11$ for $B\rightarrow\rho$ throughout the full
kinematic range.

It is then clear from Eq. (\ref{suppression}) that
the symmetry relations in Eq. (\ref{final})
are dominated by the $T_{1}$ terms as the $\alpha$ and $\beta$ terms
are suppressed by $\varepsilon$. (Near $q^{2}=t_{m}$ some relations
are further suppressed by $\varepsilon (m_{V}/m_{B})$).
Thus, the $B\rightarrow V$ hadronic form factors
satisfy the heavy-quark symmetry relations even if the quark $Q$ is
much lighter than the $b$ quark. The breakdown of the relations is a
measure of the weak binding approximation and is a small correction.

Since the effect is small and stable across the full kinematic range
we can use the model to investigate, with some confidence, the
$1/m_{Q}$ behaviour of
$\varepsilon$. This behaviour can most easily be checked near the zero
recoil part, in which the recoil effect becomes insignificant.
In Fig. \ref{curve}, we show the $m_{Q}$ dependence of
$\varepsilon (t_{m})$ with $m_{V}-m_{Q}$ fixed at
$m_{\rho}-m_{u}$. For $m_{Q}$ greater than about $0.9\,GeV$,
$\varepsilon$ falls off like power law of $1/m_{Q}$,
As shown in the figure, the curve can be approximated, above $1.25\,GeV$,
by a Taylor expansion of $\varepsilon(t_{m})$
with respect to $\langle p^{2}\rangle /m^{2}_{Q}$ using
Eq. (\ref{delta}). The leading terms in the expansion are
given by
\begin{equation}
\varepsilon(t_{m}) \approx \left(\frac{m_{V}-m_{Q}}{m_{Q}}\right)
-\frac{1}{3}\left( \frac{m_{q}}{m_{Q}+m_{q}}\right)
\frac{\langle p^{2}\rangle }{\beta^{2}_{V}}
\frac{m_{V}}{m_{Q}}\left( 1+\frac{m_{Q}}{m_{b}}\right)
-\frac{3}{8}\frac{\langle p^{2}\rangle }{m^{2}_{Q}}
\frac{m_{V}}{m_{Q}}\left( 1+\frac{7}{9}\frac{m_{Q}}{m_{b}}\right)
\,\,\, ,\label{approximation}
\end{equation}
where
\[
\langle p^{2}\rangle =
\frac{ \int d{\bf p}\phi_{V}({\bf p})\phi_{B}({\bf p})\, p^{2} }
{ \int d{\bf p}\phi_{V}({\bf p})\phi_{B}({\bf p}) }
=\frac{ 3\,\beta^{2}_{B}\beta^{2}_{V}}{\beta^{2}_{B}+\beta^{2}_{V}}
\,\,\, .
\]
The expansion parameter $\langle p^{2}\rangle$
is a stable function of $m_{Q}$ with
$\sqrt{\langle p^{2}\rangle}=428\,MeV$ for $m_{u}$ and
$\sqrt{\langle p^{2}\rangle}=502\,MeV$ for $m_{b}$.
At about $m_{Q}=0.9\,GeV$,
$\varepsilon$ has a maximum and ceases to follow the $1/m_{Q}$ power
law. It is this turn--over that stops the correction $\varepsilon$
from becoming very large for
smaller $m_{Q}$ and keeps the correction to the
symmetry relation in Eq. (\ref{final}) small for $s$ and $u$ quarks.

\vspace{.3in}
\centerline{{\bf II}. The Decays $B\rightarrow K^{\ast}\gamma$ and
$B\rightarrow\rho e\bar{\nu}$}

The branching ratio for the exclusive $B\rightarrow K^{\ast}\gamma$
to the inclusive $b\rightarrow s\gamma$ processes
can be written in terms of \(f_{1}\) and \(f_{2}\) at $q^{2}=0$,
as \cite{Alt,Desh}
\begin{eqnarray}
R(B\rightarrow K^{\ast}\gamma ) &=&
\frac{\Gamma (B\rightarrow K^{\ast}\gamma)}
{\Gamma (b\rightarrow s\gamma)}
\cong \frac{m_{b}^{3}(m_{B}^{2}-m_{K^{\ast}}^{2})^{3}}
{m_{B}^{3}(m_{b}^{2}-m_{s}^{2})^{3}}\frac{1}{2}\left[\,|f_{1}(0)|^{2}
+4|f_{2}(0)|^{2}\,\right]\,\,\, .\label{ratio}
\end{eqnarray}
Using Eq. (\ref{final}), we
can write $f_{2}(0)=(1/2)f_{1}(0)$ at $q^{2}=0$.
Although there is now only one form factor to calculate
in Eq. (\ref{ratio}), this is still a controversial model-dependent
calculation
\cite{OT,Alt,Desh,Dom,Aliev}. There is an uncertainty of about
a factor of about ten depending on the way the large
recoil of the $K^{\ast}$ is handled.

In an attempt to overcome this uncertainty,
Burdman and Donoghue \cite{BD} have
discussed a method of relating $B\rightarrow K^{\ast}\gamma$ to
the semileptonic process
$B\rightarrow \rho e \bar{\nu}$ using the static $b$-quark
limit and $SU(3)$ flavor symmetry.
Their main result is that the ratio
\begin{eqnarray}
\Gamma(B\rightarrow K^{\ast}\gamma)\left( \lim_{q^2\rightarrow 0,curve}
\frac{1}{q^2} \frac{d\Gamma (B \rightarrow \rho e \bar{\nu})}{dE_{\rho}dE_e}
\right) ^{-1}
&=&\frac{4 \pi^2}{G^{2}_{F}}\frac{|\eta|^2}{|V_{ub}|^{2}}
\frac{(m_{B}^{2}-m^{2}_{K^{\ast}})^3}{m_{B}^{4}}\,\,\, ,\label{BDeq}
\end{eqnarray}
is independent of hadronic form factors. Here, $\eta$ represents the
QCD corrections \cite{stan} to the decay
$b \rightarrow s \gamma$, and the word ``curve'' denotes the region
in the Dalitz plot for which $q^2 = 4 E_e (m_B - E_{\rho} - E_e)$.
The only uncertainty on the right hand side is that of $|V_{ub}|$ for which
\cite{PDG} $|V_{ub}|/|V_{cb}| = 0.10 \pm 0.03$.

Their method proposes to overcome the uncertainty in the calculation
at large recoil ($q^{2}=0$)
of the $B\rightarrow K^{\ast}$ form factors
by making a direct measurement of the
semileptonic decay $B\rightarrow \rho e\bar{\nu}$.
Notice that we use only the $q^2=0$ point on the ``curve"
to compare with the photonic decay in Eq. (\ref{BDeq}).
The problem with this is that the semileptonic decay
vanishes at the $q^2=0$ point on the ``curve,"
which is why this kinematic factor
is divided out in Eq. (\ref{BDeq}).
This means that experimentally there should be no events at that point
and very few in the neighborhood, making it a very difficult
measurement.

We shall overcome this
by considering instead the $q^2$-spectrum for the semileptonic
decay $B\rightarrow \rho e \bar{\nu}$. The advantage here is
that the $q^{2}$-spectrum does
not vanish at $q^{2}=0$ since we integrate over the events
from different electron energies across the Dalitz plot.
The disadvantage is that in taking the ratio we do not have the simple
cancellation of form factors, which made the previous relationship so
appealing. However, we can relate the ratio to the knowledge of
$\varepsilon$, which we have demonstrated to be a small number
anywhere in the Dalitz plot.

The differential width for $B\rightarrow\rho e\bar{\nu}$ is given by
\begin{equation}
\frac{d\Gamma (B \rightarrow \rho e \bar{\nu})}{dq^2} =
\frac{G^{2}_{F}}{12 \pi^3} |V_{ub}|^{2} |{\bf k}|^3 \Lambda_{T}\,\,\, ,
\end{equation}
where
\begin{eqnarray}
\Lambda_{T}&=& T^{2}_{1}q^{2}
+T^{2}_{2}\frac{(m^{2}_{B}-m^{2}_{\rho})^{2}}{2m^{2}_{\rho}}
\left(1+\frac{3q^{2}m^{2}_{\rho}}{m^{2}_{B}|{\bf k}|^{2}}\right)
+T^{2}_{3}\frac{2m^{2}_{B}|{\bf k}|^{2}}{m^{2}_{\rho}}\nonumber\\
&& +T_{2}T_{3}\frac{(m^{2}_{B}-m^{2}_{\rho})}{m^{2}_{\rho}}
(m^{2}_{B}-m^{2}_{\rho}-q^{2})\,\,\, .\nonumber
\end{eqnarray}
At $q^2=0$, the differential width for
$B\rightarrow \rho e \bar{\nu}$ reduces to
\begin{equation}
\left. \frac{d\Gamma (B \rightarrow \rho e \bar{\nu})}{dq^2}
\right|_{q^2=0} =
\frac{G^{2}_{F}}{192\pi^{3}} |V_{ub}|^{2}
\frac{(m^{2}_{B}-m^{2}_{\rho})^{5}}{m^{3}_{B}m^{2}_{\rho}}
|T_{2}(0) + T_{3}(0)|^2\,\,\, .
\end{equation}

If we use the symmetry relations in Eq. (\ref{final}) and
$SU(3)$ flavor symmetry in which
$T^{B\rightarrow K^{\ast}}_{1}=T^{B\rightarrow\rho}_{1}$,
we can express the ratio between $R(B\rightarrow K^{\ast}\gamma)$ and
$d\Gamma (B\rightarrow\rho e\bar{\nu})/dq^{2}$ at $q^{2}=0$ as
\begin{eqnarray}
R(B\rightarrow K^{\ast}\gamma ) &&
\left( \left. \frac{d\Gamma (B \rightarrow \rho e \bar{\nu})}{dq^2}
\right|_{q^2=0}\right) ^{-1}\nonumber\\
&=& \frac{192\pi^{3}}{G^{2}_{F}} \frac{1}{|V_{ub}|^{2}}
\frac{(m^{2}_{B}-m^{2}_{K^{\ast}})^{5}}{(m^{2}_{B}-m^{2}_{\rho})^{5}}
\frac{(m_{B}-m_{\rho})^{2}}{(m_{B}-m_{K^{\ast}})^{2}}
\frac{m^{3}_{b}}{(m^{2}_{b}-m^{2}_{s})^{3}} \,|{\cal I}|^{2}\nonumber\\
&=& 1.9\times 10^{16}\, GeV\,\cdot\, \left(\frac{0.1}{|V_{ub}/V_{cb}|}
\right) ^{2} \,|{\cal I}|^{2}\,\,\,\,\, .
\label{ours}
\end{eqnarray}

In the limit $\varepsilon \approx 0$, that is $\alpha\approx 0$
and $\beta\approx 0$ in Eq. (\ref{final}),
we have $\,{\cal I}=1\,$ in Eq. (\ref{ours}).
We estimate the correction
to ${\cal I}$ using the quark-model results for $\alpha$ and $\beta$
in Eq. (\ref{suppression}) as
\begin{eqnarray}
{\cal I} &=& \left( 1 - \frac{ (m_{B}+m_{K^{\ast}})}{2m_{B}}
\varepsilon^{B\rightarrow K^{\ast}}(0)\right)
/ \left( 1 - \frac{(m_{B}+m_{\rho})(m^{2}_{B}+m^{2}_{\rho})}
{4m^{2}_{B}m_{\rho}} \varepsilon^{B\rightarrow\rho}(0)\right)
\,\,\, .
\end{eqnarray}

Although $\varepsilon$ is a small number, the value of $\varepsilon(0)$
is sensitive to the details of the wave-function
overlap in the quark model because of the large recoil momentum.
In an extreme case with a relativistic recoil dependence
in the overlap,
we obtain $\varepsilon^{B\rightarrow K^{\ast}}(0)=0.010$
and $\varepsilon^{B\rightarrow \rho}(0)=0.043$
which gives $\,{\cal I}= 1.09\,$.
On the other hand, using a milder nonrelativistic recoil
dependence, we obtain $\varepsilon^{B\rightarrow K^{\ast}}(0) = 0.046$
and $\varepsilon^{B\rightarrow\rho}(0) = 0.087$ which gives
${\cal I}=1.18$. The uncertainty in calculating the branching ratio
$R(B\rightarrow K^{\ast}\gamma)$ due to the recoil
problem has now been reduced by an order of magnitude.

We have derived a relation between the branching ratio
$R(B\rightarrow K^{\ast}\gamma)$ and the $q^{2}$-spectrum for
$B\rightarrow\rho e\bar{\nu}$.
Since the $q^{2}$-spectrum
for $B\rightarrow\rho e\bar{\nu}$ does not vanish at $q^{2}=0$,
this reduces the uncertainty in the measurement of the semileptonic decay
in contrast to the case in Eq. (\ref{BDeq}).
Now ARGUS has given the result \cite{argus} of
$BR(B^{-}\rightarrow \rho^{0}l\bar{\nu})=(11.3\pm 3.6\pm 2.7)
\times 10^{-4}$, and by isospin symmetry,
$\Gamma (\bar{B}^{0}\rightarrow \rho^{+}l\bar{\nu})=
2 \Gamma (B^{-}\rightarrow \rho^{0}l\bar{\nu})$.
This allows us to estimate
$d\Gamma (B\rightarrow\rho e\bar{\nu})/dq^{2}$ at $q^{2}=0$
to be about $10^{-17}\, GeV^{-1}$. Eq. (\ref{ours}) then gives
$R(B\rightarrow K^{\ast}\gamma)$ about $10^{-1}$,
which is quantitatively correct.
A direct measurement of
$d\Gamma (B\rightarrow\rho e\bar{\nu})/dq^{2}$ at $q^{2}=0$ can
therefore provide reliable information for
$R(B\rightarrow K^{\ast}\gamma)$.

\vspace{.3in}
\centerline{ {\bf Acknowledgment}}

This work was supported by the Natural Sciences and Engineering
Council of Canada and by the National Sciences Council of the
Republic of China.

\figure{The $m_{Q}$ dependence of $\varepsilon(t_{m})$ with
$m_{V}-m_{Q}$ fixed at $m_{\rho}-m_{u}$. The solid line is the
exact numerical result of $\varepsilon (t_{m})$ using Eq. (\ref{delta}).
The dotted line is a Taylor expansion of $\varepsilon (t_{m})$ with
respect to $\langle p^{2}\rangle /m^{2}_{Q}$.
\label{curve}}

\epsffile[100 400 438 632]{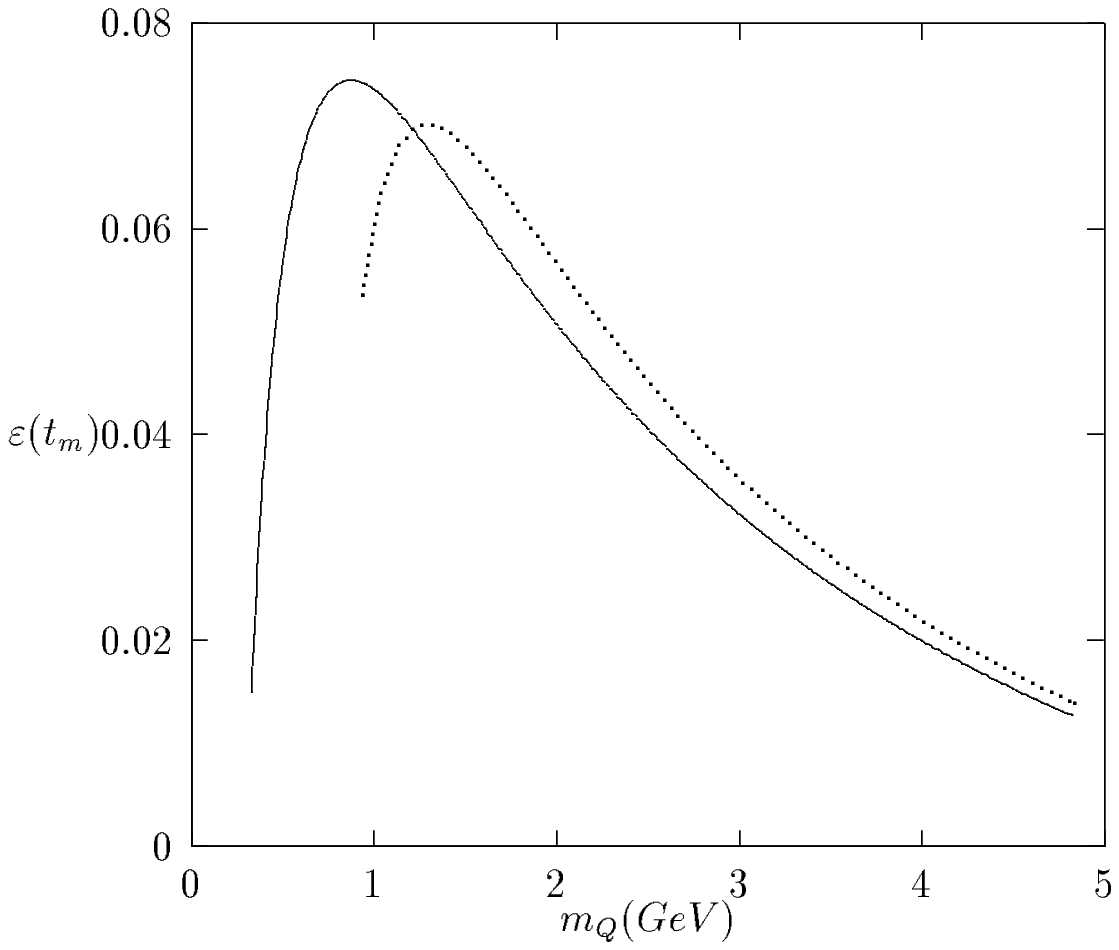}
\end{document}